\documentclass[10pt,preprint]{aastex}%
\begin{document}

\title{The Spectral Energy Distribution of the High-Z Blazar Q0906+6930}
\author{Roger W. Romani}
\affil{Department of Physics, Stanford University, Stanford, CA 94305}
\email{rwr@astro.stanford.edu}

\begin{abstract}
	We describe further observations of QSO J0906+6930, a z=5.48 blazar
likely to be detected in $\gamma$-rays. New radio and X-ray data place
significant constraints on any kpc-scale extension of the VLBA-detected jet.
Improved optical spectroscopy detects absorption from an intervening galaxy at
z=1.849 and raise the possibility that this distant, bright source is
lensed. We combine the new data into an improved SED for the blazar
core and comment on the Compton keV-GeV flux component.
\end{abstract}


\section{Introduction}

	Q0906+6930 (=GB6 J0906+6930=SRM J090630.74+693030.8)
was discovered in a survey of radio-bright, flat spectrum
sources chosen to be like the EGRET blazars \citep{srm03,set05}.  Follow-up
observations confirmed a large $z\approx 5.5$ redshift and found evidence
for a very compact pc-scale jet with the VLBA \citep{ret04}. The source is
quite radio-loud with an $R = f_{\rm 5\,GHz}/f_{\rm 440nm}({\rm rest}) 
\approx 500-1000$, supporting the blazar interpretation.  These data 
showed several peculiar properties. For example, the radio spectrum appeared 
to steepen above 10\,GHz, while the jet component still appeared to be inverted.
Also, the source shows a relatively large $1350$\AA\, continuum flux and 
large kinematic widths for the emission lines. These suggest a large 
$\sim 2 \times 10^9M_\odot$ black hole mass [note that an error in the 
$\lambda F_\lambda$ luminosity quoted in \citet{ret04} implied 
$M\ge 10^{10}M_\odot$].
While this source was, at best, a weak background enhancement in the 
{\it EGRET} survey, the prospects for detection with GLAST seem quite strong. 
This is particularly interesting as observations of a cut-off in the blazar
spectrum above $\sim 10$\,GeV can be used to probe absorption by
light produced at the peak of star formation \citep{crr04}.

	We report here on further observations of Q0906+6930, which support
its identification as a high-z blazar, test the nature of the jet component
and probe it's status as a high mass, high luminosity source.	

\section{New Observations}

	Several radio-loud QSO at high redshift show kpc-scale jets
so our first objective was to constrain such arcsecond-scale emission.
At the blazar redshift, $1^{\prime\prime}= 6.19$\,kpc (for an 
$\Omega_m= 0.24$, $H_0= 73$\,km/s/Mpc flat cosmology, assumed throughout; 
Spergel et al. 2006). Perhaps the most convincing case of such jet emission
is still that of QSO GB 1508+5714 at z=4.3, which is
detected both in the X-ray \citep{sie03} and radio \citep{che04} bands.
Its large observed $f_X/f_R \sim 10^2$ supports a scenario where the 
jet electrons Compton up-scatter CMB seed photons, which have a large
increase in energy density, $u \propto (1+z)^4$ at high redshift \citep{sch02}.

\subsection{X-ray imaging}

	We obtained a 30\,ks {\it Chandra} observation of the blazar
on 2005 July 1 positioned at the best focus of the ACIS-S3 
backside-illuminated chip. The observations were standard full-frame
3.2s TE exposures in very faint (VF) mode.
No strong background flaring was seen during this observation so that 
all of the data could be included in the analysis.
The data were processed using standard algorithms in CIAO version
3.2 and CALDB version 3.0.0. 
The blazar was clearly detected, providing $\sim 435$ counts, at 
a position coincident with the radio core.  With this modest flux, 
pile-up is under 1\%, and so we can ignore it in the subsequent analysis.  
The X-ray image is consistent with a PSF computed for the blazar 
spectrum; the largest excess (at $\sim 1.5^{\prime\prime}$, position
angle $\sim 115^\circ$) contains 4 photons and is not significant.
Thus, on $\sim 1-10$ arcsecond scales, any
extended (jet) emission must have $\le 1$\% of the core flux. 
A number of coronal emission field stars and some unidentified 
objects (presumably background AGN) were detected in this integration, but
none showed notable peculiarities, so we do not discuss these further.

\begin{figure}[!h]
\plottwo{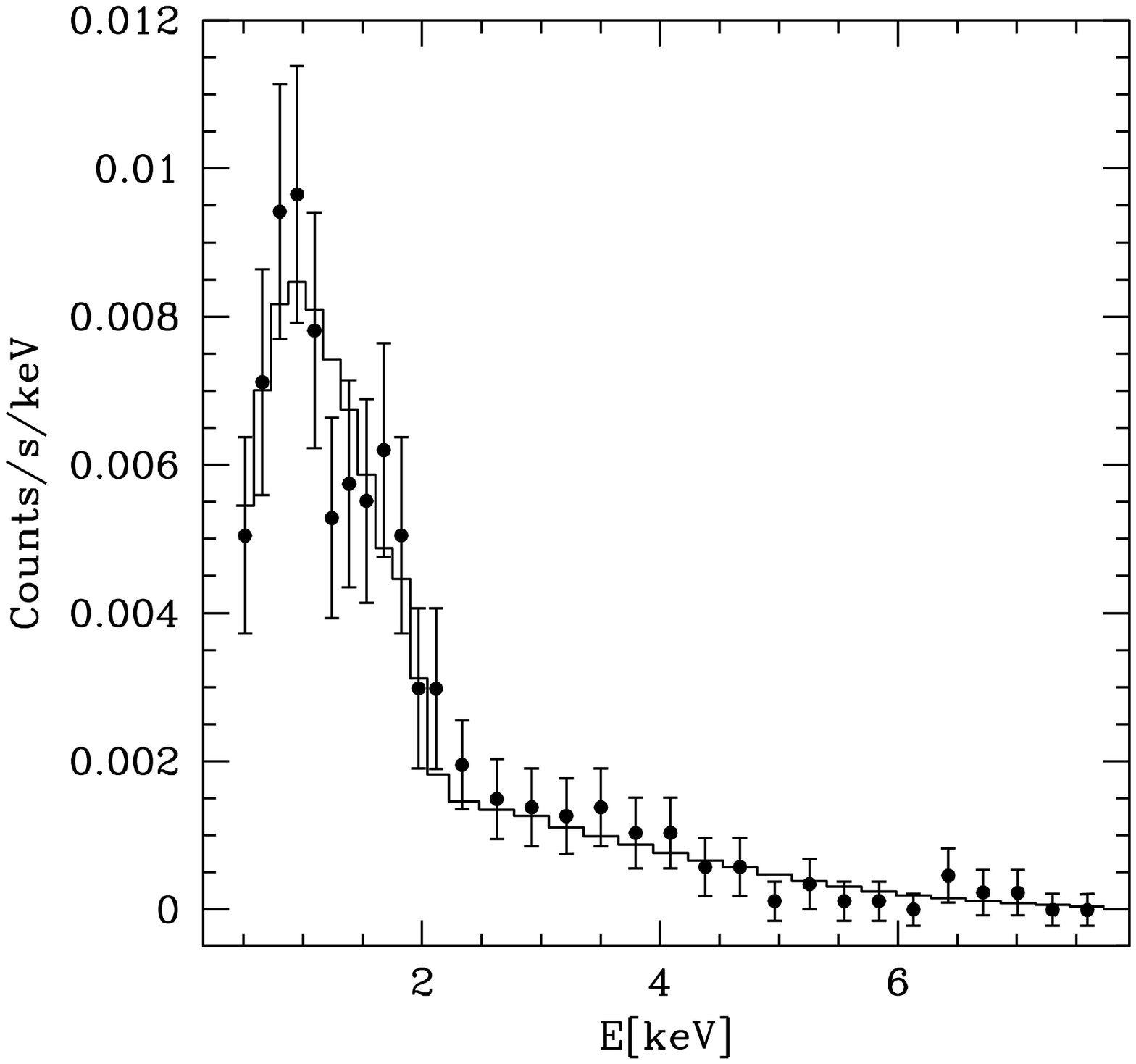}{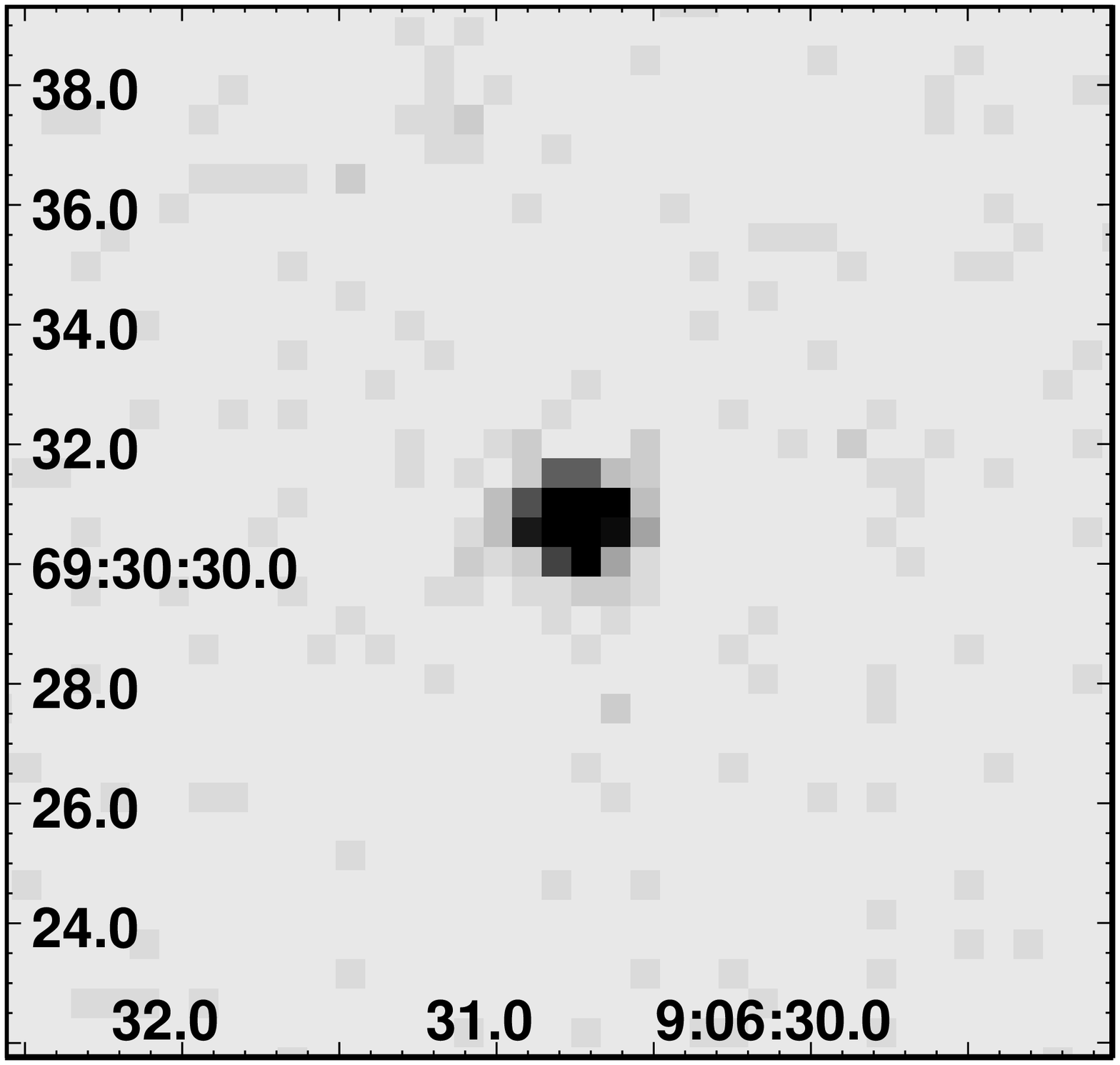}
\figcaption{\small
{\it CXO} ACIS-S3 spectrum of Q0906+6930, with the best-fit absorbed
power-law model. The second panel shows the 0.5-8\,keV {\it CXO} image.
}
\end{figure}

The X-ray spectrum (figure 1) is adequately described by a simple 
absorbed power-law,
with a low absorption $N_H = 5.4\pm2 \times 10^{20} {\rm cm^{-2}}$,
a photon index $\Gamma = 1.6 \pm 0.1$ and an unabsorbed flux
$f_X (0.2-10\,{\rm keV})) = 1.47\pm 0.25 \times 10^{-13} 
{\rm erg\,cm^{-2}\,s^{-1}}$. 
Although the data show a small excess to the best fit
power-law model around 1\,keV, an attempt to fit a red-shifted Fe line
produced no significant detection. The upper limit
to the line flux is $\sim 1.5 \times 10^{-15}{\rm erg\,cm^{-2}\,s^{-1}}$,
or $\le 1$\% of the continuum flux. These spectral parameters are quite
typical of lower redshift blazars. The measured absorption is consistent
with the Galactic column in this direction, $\sim 4.4 \times  10^{20} {\rm cm^{-2}}$ from
the COLDEN tool. We can also compare the X-ray properties with
those of a set of radio-loud high-redshift QSO recently reported by 
\citet{lopet06}. Our observed $\Gamma$ and inferred 
$\alpha_{OX} [250\,{\rm nm/2\,keV}] (\rm rest) \approx -1.3$, are in excellent
agreement with the objects measured in this paper. We conclude that this
is a fairly typical radio-bright quasar at high redshift.

\subsection{Radio imaging}

	To search for a larger scale counterpart of the VLBI jet reported
in \citet{ret04}, we also obtained VLA images with the A array at 
1.4-, 4.8- and 8.4\,GHz on 2004, October 7. The on-source times were 
10m, 30min and 50min
respectively. These data were calibrated and mapped using
standard AIPS routines and were found to be adequately described by
a simple point source model. The final map noise was within a factor
of $\sim 2$ of the thermal noise in each case. Measuring the rms background
in an annulus between the beam size and $\sim 5^{\prime\prime}$,
we obtain limits on any jet component (Table 1). The largest 
map fluxes in the searched regions are consistent with statistical
fluctuations for these rms vales, so we can take upper limits on
the jet flux along the position angle of the mas jet to be $\sim 2\times$
the local rms, while the limits on any extended component are about
two times larger.  For the X-band image this search extends within 
$\sim 0.4^{\prime\prime}$ of the blazar core, while for Q-band in
the B-array, we searched down to $0.1^{\prime\prime}$.

	Since no extended emission is seen in the A-array data,
we used follow-on B-array observations on 2005 March 04 to obtain a 
six frequency, single epoch radio spectrum. The on-source total scan times
are listed in Table 1. The data were calibrated by observations of
0542+498 (=3C 147) at each frequency. For all measurements, the statistical
error in the flux density estimate was small. We add this in quadrature
to the flux scale uncertainties described in the VLA calibrator manual 
(2\% at L, C and X and 5\% at
U, K and Q) to obtain the flux density errors in table 1. For 
K and Q residual pointing errors and uncorrected atmospheric absorption
may contribute additional error.
In all cases a single unresolved component fit provided
a satisfactory description of the Q0906+6930 data. In Table 1, we also
give RMS fluctuation levels in an annulus about the point source. The local rms 
limits to extended flux for the three lower frequency bands are from the 
deeper, higher-resolution A-array images.

\begin{table}[!ht]
\caption{Radio Core Fluxes and Jet Bounds}
\medskip
\tabletypesize=\small
\begin{tabular}{llllll}
Band & Freq(GHz) &On-Source time (min)& Core Flux (mJy) & Error (mJy) & Jet Annulus RMS (mJy)\\
\hline
L & 1.44 & 8 & 92  & 2.0 & 0.10$^\dagger$\\  
C & 4.88 & 8 & 114  & 2.4 & 0.05$^\dagger$\\ 
X & 8.44 & 12 & 136  & 3.1 & 0.13$^\dagger$\\ 
U & 14.96 & 24 & 129  & 6.6 & 0.23\\
K & 22.49 & 15 & 83  & 4.2 & 0.22\\
Q & 43.33 & 30 & 43  & 2.2 & 0.26\\
\end{tabular}
\leftline{$^\dagger$ RMS from 2004/10/04 A-array maps}
\label{radio_flux}
\end{table}

\subsection{Optical Spectroscopy and Imaging}

	To further investigate the absorbed optical spectrum reported
in \citet{ret04}, we obtained higher resolution spectra using the
Hobby$\bullet$Eberly telescope \citep[HET;][]{ram98}
Marcario Low Resolution Spectrograph \citep[LRS;][]{hill98} using
the G3 VPH grating, which provides 1.88\AA/pixel dispersion over the range
$\lambda\lambda$6250\AA\, - 9100\AA. The integrations used a $1.5^{\prime\prime}$\
slit, giving an effective resolution of
$5.5$\AA. Standard queue observations were made on 2005 February 9 and 14
and 2005 May 9-10; conditions were generally sub-par with poor transparency
or variable seeing. After standard IRAF processing and combination using 
the relative S/N, seven exposures totaling 8600s
of integration were assembled to produce the spectrum in figure 2.
Here the new G3 spectrum is plotted with a heavy line over the lower resolution
G1 spectrum of \citet{ret04}. The excellent correspondence between the
fine structure in these spectra, especially in the Ly$\alpha$ forest region,
attests to the stable calibration. The combined spectrum has a S/N per 
resolution element of $\sim 16$ in the unabsorbed continuum.

\begin{figure}
\plotfiddle{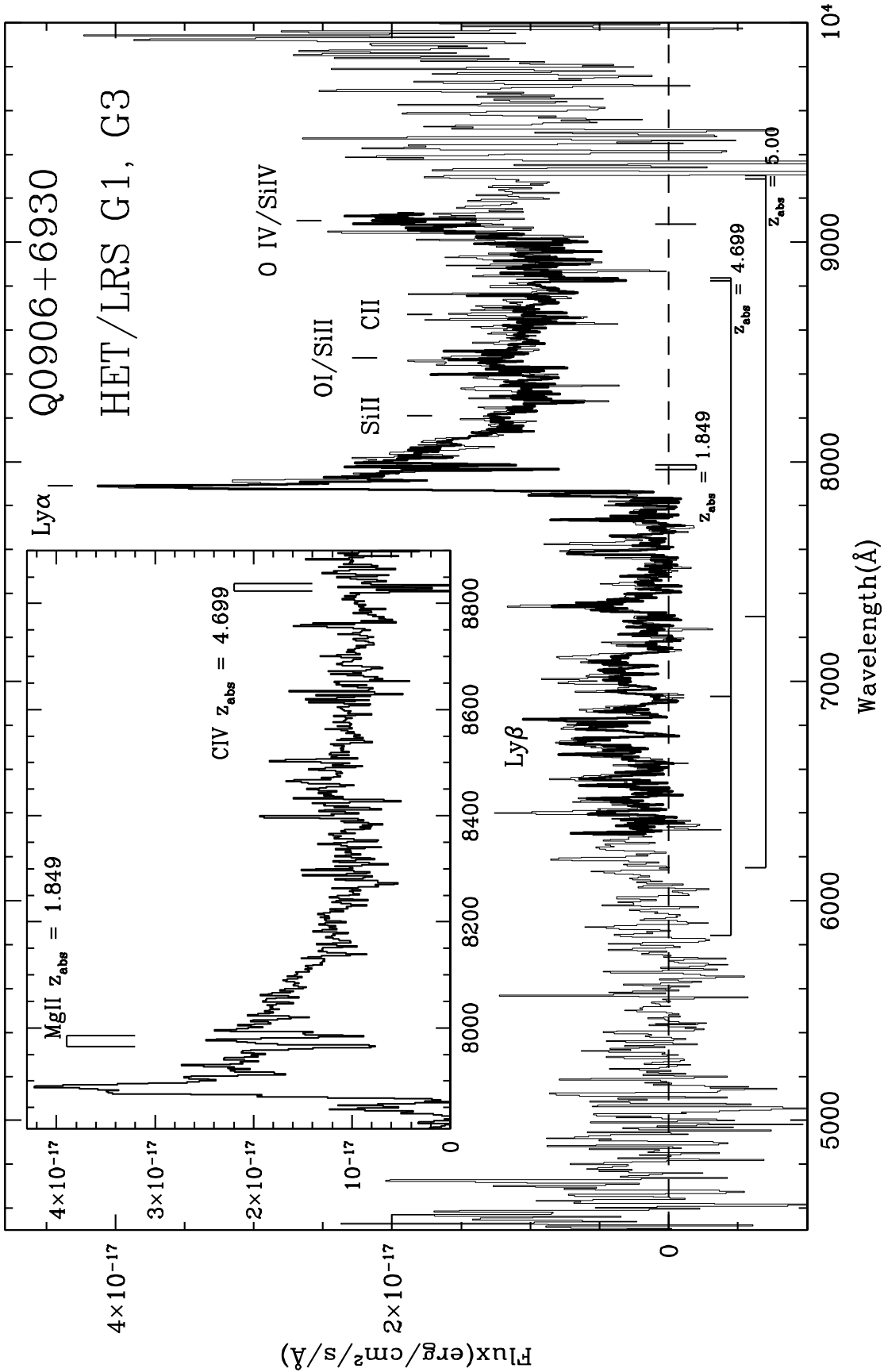}{.0truecm}{90}{650}{420}{0}{00}
\figcaption{\small
HET/LRS G3 spectrum of Q0906+6930, over plotted on lower resolution
data with a wider wavelength coverage. The inset shows the G3 data alone, 
with two metal line absorption systems indicated.
}
\end{figure}

	These higher resolution data allow us to check the reality of
absorption-line systems suggested in the G1 spectrum. Two metal line
systems can be seen in the inset of figure 2, which shows the spectrum redward
of Ly$\alpha$. In particular, we confirm a likely C IV doublet at $z=4.698$, 
with a rest equivalent width (REW) for each component of $W_0=0.6\pm 0.13$\AA. 
This system has an associated, apparently damped, Ly$\alpha$ absorption
and a minimum at Ly$\beta$. More interesting is an apparent strong Mg II
absorption at $z=1.849$ which was confused by possible N\,V emission in the
lower resolution data. The REW of the 2795\AA\, line is $1.9\pm0.07$\AA\,
and the line spacing and doublet ratio $R=1.2$ support this identification.
In our restricted spectral range no strong confirming line is expected.

	The presence of intervening absorption line systems raises the
possibility that the blazar is magnified by gravitational
lensing, as discussed below. To constrain this hypothesis, we examined
the limited imaging data. During HET queue spectroscopy, 10-30s
$\sim$white light set-up images are generally recorded through a 515nm 
long-pass filter. Six of these images had usable ($\le 1.7^{\prime\prime}$)
image quality; these were S/N weighted and combined to produce 
a final image with $1.6^{\prime\prime}$ FWHM. This was used to
get an approximate limit on any extended source within $5^{\prime\prime}$
of Q0907+6930 of $R\sim23.1$. Imaging reconnaissance was also obtained
using the MDM 2.4m RETROCAM on 2006 February 4, courtesy of J. Marshall 
and D. DePoy. Here $6\times 300$s $r$ images ($\sim 1.3^{\prime\prime}$ FHWM)
were combined. Although Ly$\alpha$ forest absorption should suppress
the quasar relative to the foreground galaxy in the $r$ band, in the
combined image the blazar was again stellar, with an apparent magnitude
of $r=21.9$. We derive an upper limit of $r>23.3$ on any 
nearby galactic companion.

\section{Spectral Energy Distribution}

	We have assembled these new flux measurements into a (non-instantaneous)
spectral energy distribution (SED), figure 3. The plot shows the 04/03/05
simultaneous radio fluxes. We do not see any strong evidence of radio 
variability, but it should be noted that the X-ray flux (shown with a dashed 
lines for the continuum uncertainty range) was observed $\sim 100$d later.
The optical spectroscopy measurements span several years, although again we
do not see variability. With a rather small range of unabsorbed continuum
available, the optical spectral index is quite uncertain (dashed lines, extended a
a decade for visibility). This situation would be greatly helped by near-IR
spectroscopy. Finally, the {\it EGRET} upper limit is, of course, from
10-15 years prior to the current observations, and blazars are known to be
highly variable in the $\gamma$-ray band. The low significance detection during
one viewing period \citep{ret04} should be seen as consistent with the 
mission-averaged upper limit.

	As in \citet{ret04} we compare the new SED with that of the
well-studied EGRET blazar 3C 279 [inset: SED from period P5b \citep{har01},
red-shifted to z=5.48], and with synchro-Compton models. The Q0906+6930
spectrum is sparse, but notably shows a significantly ($\ge 3\times$) more
luminous optical spectrum. In contrast to 3C 279 the optical emission shows
high equivalent-width lines from the broad-line region (BLR),
and is thus not dominated by synchrotron continuum.  For the Compton components 
of the SED, our new X-ray detection confirms a rising $\nu F_\nu$
spectrum, likely up-scattered by the synchrotron-emitting electrons. 
This is somewhat ($\sim 50$\%) fainter than the 3C 279 X-ray luminosity, 
which is reasonable in light of the lower radio synchrotron luminosity. 
However, the possible GeV detection would require an external Compton component 
$\sim 30\times$ brighter than that of 3C 279.

   By computing representative synchro-Compton models, we see that the brighter
disk emission and a luminous $\gamma$-ray Compton component
may be related. In the inset we show a model for 3C 279 computed
from a code kindly supplied by M. B\"ottcher. This code computes synchrotron
and Compton emission in an evolving cylindrical jet outflow. It also models a host
accretion disk and computes the Comptonized flux from these soft disk photons,
both directly from the disk and disk photons scattered from the broad line 
region. The code is described in \citet{boet97} and \citet{bb00}. To generate the 3C 279
model, we use explicitly the parameters listed in \citet{har01} for this
epoch. The `jet' is a $6\times 10^{16}$\,cm radius blob injected 0.025\,pc 
from the $1.5\times 10^8M_\odot$ hole, moving cylindrically at $2^\circ$ 
to the line of sight with
bulk $\Gamma=13$. The jet contains 35${\rm cm^{-3}}$ each of $e^\pm$,
with a power-law energy distribution of index $p=3.0$, extending from
$\gamma_1=6\times 10^2$ to $\gamma_2=1\times 10^5$ and 1.5\,G of
magnetic field. The hole is surrounded
by an accretion disk of luminosity $10^{46}{\rm erg/s}$ and a broad-line
region extending from 0.1-0.4\,pc. To reproduce the model shown in
\citep{har01}, it was necessary to increase the BLR Thomson depth to $\tau=0.06$.
These parameters gave the total SED curve shown in the inset.

    We know that Q0906+6930 has a higher mass hole ($\sim 2 \times 10^9M_\odot$)
and a brighter accretion disk ($L_D \ge 3 \times 10^{46}{\rm erg/s}$). We
made these two changes to the parameters and re-computed the SED, as shown
in the main part of Figure 3. Under the full SED we show the individual
components: (from low to high frequency) the jet synchrotron emission,
the disk thermal emission, the jet self-Compton emission, the external Compton
emission from the disk and the Comptonized emission from disk photons 
scattered off the broad-line
region. The spectra were computed at $z=0.538$ and then the 3C 279 data and
the two models were red-shifted to z=5.48 using concordance flat cosmology 
parameters.  These models illustrate that with a higher mass hole and
a brighter, cooler disk, brighter external Compton ($\gamma$-ray) emission
can be expected. Of course, such models have many parameters and fits to more
densely sampled SEDs in a variety of flare states are needed for any predictive 
power.

	A standard definition of radio-loudness is the ratio 
$R=f_{\rm 5\,GHz}/f_{\rm 440nm}({\rm rest})$ \citep{ket89}. Although we do not 
have direct
observations at $770$\,MHz or $2.9\mu$, we can extrapolate the observed
optical and radio spectra to estimate $R\approx 540-970$ where the range
is due to the rather uncertain optical continuum spectrum.

       We should note that both the 3C 279 and the Q0906+6930 models
safely under-predict the radio spectrum. This is because the model
simulates only the base of the jet emission, omitting larger
scales where cooling, expansion and injection of additional $e^\pm$ populations
produce the cm-wavelength radio components. In particular, the
local maximum in the Q0906+6930 radio spectrum at $\sim 15$GHz ($\sim 100$GHz 
rest-frame) suggests a distinct radio jet component. It should be noted that
the 43\,GHz flux measured here (on kpc scales) is in good agreement with the 
{\it VLBA} core measurement. Thus at most 10\% percent of the radio flux lies 
in the $\sim 5-10$\,pc resolved VLBA jet and the cm-wavelength radio component
is produced at sub-pc scales in the unresolved VLBA core.
It would be useful to check if the SED rises
toward a synchrotron peak in the far-IR/sub-mm region as expected. 
IR photometry of the blazar will also be helpful in constraining the 
level of the synchrotron SED and confirming the dominance of disk emission
in the optical.

\begin{figure}[h!]
\plotone{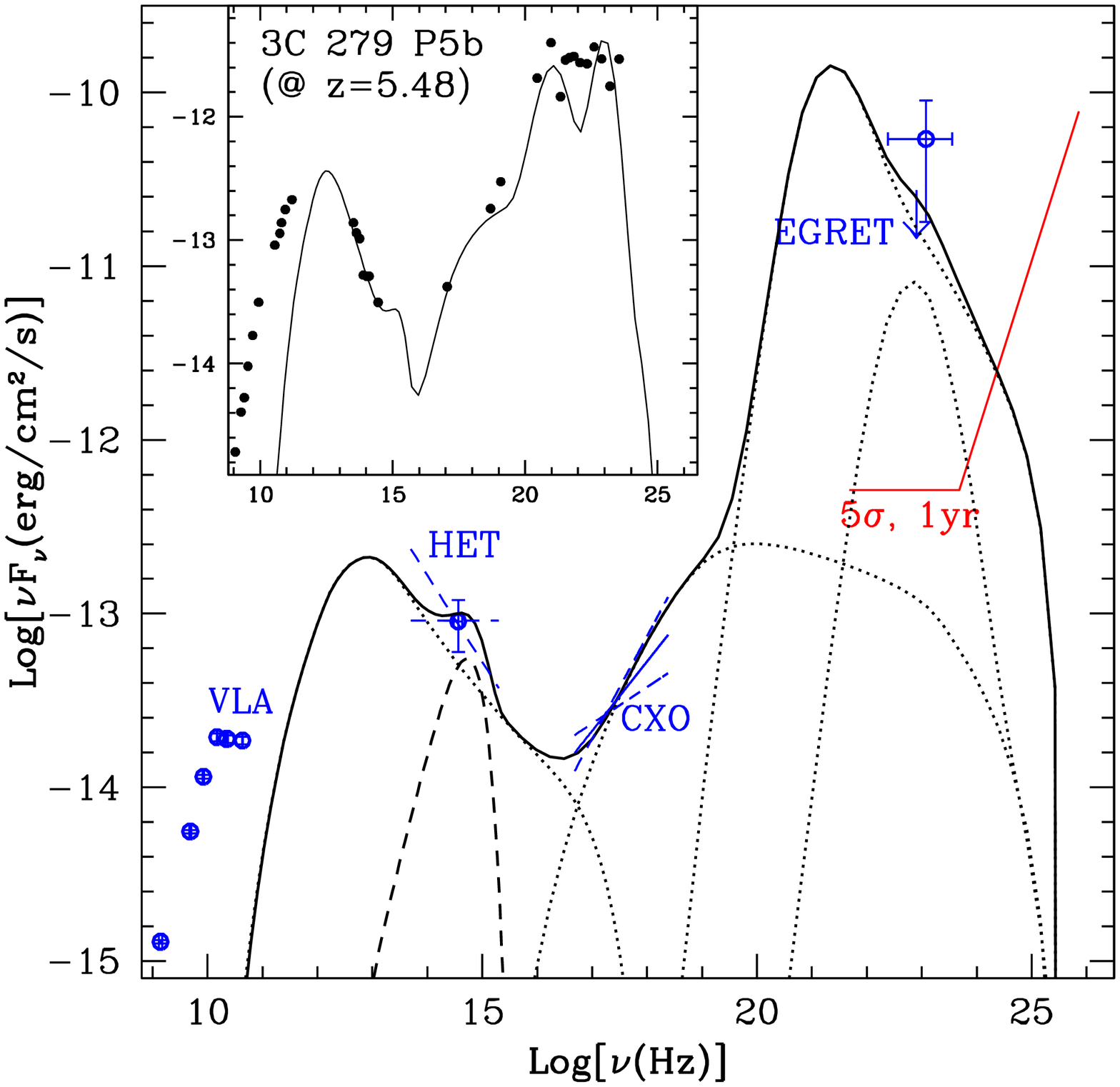}
\figcaption{\small Multiwavelength spectral energy distribution for Q0906+6930,
from non-simultaneous data (open circles).
The optical continuum flux is plotted with a line (short dash) extended to
show the slope. The X-ray spectrum (solid line) shows our {\it CXO} 
continuum fit. For these
two components dashed lines show the range of spectral index uncertainty.
For the $\gamma$-ray band we show the brightest epoch {\it EGRET} flux,
the {\it EGRET} mission-averaged upper limit and the {\it GLAST} sensitivity
of the first year sky survey. A synchro-Compton model is plotted (see text).
The individual component curves (synchrotron emission, synchrotron self-Compton
emission, disk emission [dashed line] and Comptonized emission directly
from the disk and after scattering in the broad line region) are shown.
For comparison, the inset shows the SED of 3C 279 during period P5b \citet{har01},
shifted to z=5.48, along with its synchro-Compton model (see text).
}
\end{figure}

\section{Conclusions}

	The new data generally confirm the SED picture described in
\citet{ret04} and show that at least the X-ray Compton component is present in
this blazar. The lack of any resolved kpc-scale jet component suggests
that all emission observed is from the core (or a compact pc-scale jet).
Also, this does not support the \citet{sch02} picture where large-scale
X-ray jets are produced by up-scatter of the CMB. In this model, the 
$\propto (1+z)^4$ increase of the CMB energy density counters the 
decreasing jet surface brightness; at $z>4$ jet emission is thus expected
to exceed
core emission. With $\le 1$\% of the core X-ray flux in a resolvable jets,
our results extend those of \citet{lopet06}, suggesting core photons,
rather than CMB photons dominate the flux up-scattered as Compton
X-ray emission.

One peculiarity noted in \citet{ret04} was the relatively large optical and radio
flux; we noted the possibility of lensing of the QSO core. This should be
re-visited, since the strong intervening metal line absorbers suggest 
galactic-scale
masses along the line-of-sight. The most important is the MgII system at
z=1.849, which at  $W_0=1.9$\AA, likely represents a classical $\sim L_\ast$
spiral galaxy \citep{chet05} which should be a Lyman Limit System (LSS)
absorber, within an impact parameter
$$
R = 38 h^{-1}(L_K/L_K^\ast)^{0.15} {\rm kpc},
$$
for an $L_\ast$ galaxy at the absorber redshift, i.e. within $6.2^{\prime\prime}$ 
for our assumed cosmology. Our limited direct imaging does not detect such a 
galaxy, but the results are not very constraining. At this $z$, the cosmic age is 
$\sim 3.6$\,Gyr and an $L_\ast$ Sa formed at z=5 in our cosmology would have 
$r\sim 23.9$ for a \citet{bc93} model,
i.e. about $2\times$ fainter than our imaging $r\sim 23.3$ limit. Sbc types
would be even fainter.

	Can such a galaxy lens our blazar? The expected image angular scale
for a lens mass $M$ is
$$
\theta_0 = \left [ 4 D_{LS}G\,M/(D_S D_L c^2)\right ]^{1/2}
$$
where one uses angular diameter distances and $D_{LS} = D_s-(1+z_L)D_L/(1+z_s)$. 
The geometry is favorable for lensing and the lens scale is then
$1.35 \times 10^{-6}M^{1/2}$arcsec with $M$ in $M_\odot$. We see no evidence for a 
double image of the blazar on arcsecond scales, either in the optical or 
the radio, so this precludes classical macro lensing by an $L_\ast$ galaxy.
Indeed, the core is unresolved in our VLBI maps, where the half-power beam width
at 43\,GHz was 0.55$\times 0.30$\,mas, so we can place an upper limit on
an effective lens mass of $M\le 1.4 \times 10^5 M_\odot$. The lack of obvious
variability on year time-scales suggests a not very constraining lower
mass limit of
$M\ge 10^{-3} v_{300}^2 M_\odot$ for a lens galaxy velocity dispersion
of $\sim 300\,v_{300}$\,km/s. We may infer a slightly tighter limit if
we note that the large BLR equivalent widths imply that the broad line
region must be lensed along with the core. Naively applying the continuum
BLR radius correlation of \citet{kasp05} to our optical continuum luminosity
$\lambda F_\lambda |_{5100\AA} \approx 3 \times 10^{46} {\rm erg/s}$
we conclude
$$
R_{BLR} \approx 5.8 \times 10^{16} 
\left ( \lambda F_\lambda |_{5100\AA}/10^{44}{\rm erg/s} \right )^{0.69}
\approx 2.9 \times 10^{18}{\rm cm}.
$$
This in turn implies a minimum lens mass $M\ge 10^{-2}M_\odot$. We infer
that macro-lensing does not amplify Q0906+6930, but that micro- or milli-lensing
are possible. On the other hand, it should be noted that the probability
of an unrelated intervening galaxy is not small. Since $dN/dz(W_0>1.9\AA) \approx
0.13$ over the redshift range covered by our G3 spectrum \citep{net05}, we 
expect 0.06 such absorbers. Detection of one MgII systems is thus not very improbable 
and so we cannot infer from this that lensing amplification has brightened 
Q0906+6930.

	We conclude by re-iterating that our new SED measurements
support the picture of Q0906+6930 as a blazar with a bright $\gamma$-ray
Compton component emitting at z=5.48. This Compton radiation must
traverse the peak of star formation at $z\approx 2-3$, where 
optical/UV emission can attenuate the $>$10\,GeV photons, just as the IR
background can attenuate the TeV emission of blazars at lower $z$,
e.g. \citet{dka05} and references therein. 
Of course, there is also
attenuation from the host photon field and, in principle, from the intervening
z=1.89 galaxy. These contributions will make it difficult to extract 
constraints on the extragalactic background photon field from this (or any one)
object. However, statistical studies of high red-shift blazars with GLAST
\citep{crr04}, should still be able to extract global constraints on the
extragalactic background light and its evolution.

\acknowledgments

{\small
We thank the referee for a careful reading. We also thank many
colleagues for assistance with the collection of the SED data: in the radio
Lincoln Greenhill and Greg Taylor, in the optical Steve Healey and in the
X-ray Stephen Ng. MDM images were kindly obtained by Darren DePoy and
Jennifer Marshall.
This work was supported in part by NASA grant G05-6101X issued by the Chandra 
X-ray Observatory Center, which is operated by the Smithsonian Astrophysical 
Observatory for and on behalf of the National Aeronautics Space Administration 
under contract NAS8-03060. 

The Hobby-Eberly Telescope (HET) is a joint project of the University 
of Texas at Austin, the Pennsylvania State University, Stanford 
University, Ludwig-Maximilians-Universit\"at M\"unchen, and 
Georg-August-Universit\"at G\"ottingen. The HET is named in honor of 
its principal benefactors, William P. Hobby and Robert E. Eberly.
The Marcario Low Resolution Spectrograph is named for Mike Marcario 
of High Lonesome Optics who fabricated several optics for the instrument 
but died before its completion. The LRS is a joint project of the 
Hobby-Eberly Telescope partnership and the Instituto de Astronomia de 
la Universidad Nacional Autonoma de Mexico.
The National Radio Astronomy Observatory is a facility of the 
National Science Foundation operated under cooperative agreement
by Associated Universities, Inc.
}

\end{document}